# Nanoporous AuPt and AuPtAg alloy co-catalysts formed by dewetting-dealloying on ordered TiO$_2$ nanotube surface lead to significantly enhanced photocatalytic H$_2$ generation

Nhat Truong Nguyen,[a] Selda Ozkan,[a] Ondrej Tomanec,[b] Xuemei Zhou,[a] Radek Zboril,[b] Patrik Schmuki[a,b,c]*

[a] Department of Materials Science and Engineering WW4-LKO, University of Erlangen-Nuremberg, Martensstrasse 7, D-91058 Erlangen, Germany.

[b] Regional Centre of Advanced Technologies and Materials, Department of Physical Chemistry, Faculty of Science, Palacky University, Slechtitelu 11, 783 71 Olomouc, Czech Republic.

[c] Chemistry Department, Faculty of Sciences, King Abdulaziz University, 80203 Jeddah, Saudi Arabia.

*Corresponding author. Email: schmuki@ww.uni-erlangen.de

Link to the published article:

https://pubs.rsc.org/en/content/articlehtml/2018/ta/c8ta04495c

**Abstract**

Effective co-catalysts are of key importance for photocatalytic $H_2$ generation from aqueous environments. An attractive co-catalyst candidate are AuPt (metastable) alloys due to the synergistic electronic and chemical interaction of the constituents in the charge transfer and $H_2$ evolution process. Here we introduce the fabrication of AuPt alloy nanoparticles with nanoporosity (pore size of 2-5 nm) fabricated on spaced $TiO_2$ nanotubes. By dewetting a layered AgAuPt coating, we form AuPtAg alloy nanoparticles. From these alloys, Ag can selectively be dissolved leading to the desired nanoporous AuPt alloy particles with diameter in the range of 10-70 nm deposited as a gradient on the $TiO_2$ nanotubes. A significant enhancement of photocatalytic $H_2$ generation is obtained compared to the same loading of monometallic or nonporous alloy. The nanoporous AuPt particles provide not only a large surface area to volume ratio (and are thus more effective) but also show the intrinsic synergy of a AuPt alloy for $H_2$ generation.

Since the first study of Fujishima and Honda on the ability to "split" water in 1972,[1] $TiO_2$ has become the most investigated semiconductor for photocatalysis due to its photostability and versatility for diverse photocatalytic applications such as pollution degradation, synthesis of organics, $CO_2$ reduction and $H_2$ generation.[2–9] When $TiO_2$ is irradiated with sufficiently high-energy light (>3.2 eV for anatase), electrons and holes are generated. These charge carriers may react with redox couples in the environment. In $TiO_2$, the band positions are suitable to allow conduction band electrons to reduce water to $H_2$ and valence band holes to oxidize water or organic solvents.[10,11]

Most frequently, $TiO_2$ nanoparticles (NPs) in the form of suspensions are used as photocatalyst owing to their high specific surface area and short diffusion distance for excited carriers to reach the surface. However, under these open-circuit conditions, the photocatalytic $H_2$ production efficiency of $TiO_2$ is limited by a relatively high electron-hole recombination rate. To overcome this, a co-catalyst, such as a noble metal NP (usually Au, Pt and Pd), is decorated on the $TiO_2$ surface.[12,13] Specifically, these metal particles can act as electron transfer mediator and $H_2$ recombination catalyst. The size, quantity and distribution of co-catalyst have been reported to strongly affect the photocatalytic performance.[13,14] Therefore, finding appropriate techniques granting an optimized co-catalyst decoration on $TiO_2$ has become essential. Nevertheless, using a $TiO_2$ NP system allows only for a very limited control of the density and placement of the noble metal decoration.

In contrast, 1-dimentional $TiO_2$ nanotubes (NTs) allow the fine control over co-catalyst placement.[15,16] Moreover, these tubes have attracted considerable interest in photocatalytic $H_2$ generation because they possess beneficial charge transport and charge separation properties.[2,17,18] These $TiO_2$ NTs are grown by electrochemical anodization from a metallic Ti substrate, they grow vertically aligned and the dimensions of the NTs can easily be controlled by electrochemical parameters.[17]

The efficiency of photocatalytic $H_2$ generation strongly depends on placement and the type of co-catalyst. Among several noble metals, Pt is the most effective co-catalyst, followed by Au, Pd and Ag, because Pt has the highest work function among the four metals and additionally catalyze the hydrogen formation reaction ($2H^0 \rightarrow H_2$).[12,19–22] A particularly interesting case is the establishment of bimetallic co-catalysts. AuPt alloys have shown synergy of the involved elements, mainly in the use as a catalyst for the electrochemical methanol oxidation or oxygen reduction and in some cases to catalyze the $H_2$ evolution reaction.[23–26] In general, Au and Pt are immiscible in the bulk, but they can form alloys at the nanoscale.[24,27] In a few cases, the use of alloyed AuPt for photocatalytic applications has been reported.[28,29] The better performance of combined Au and Pt co-catalysts, compared to a single metal decoration, has been explained by the presence of surface plasmon resonance and $Ti^{3+}$ sites/$O^{2-}$ vacancies,[28,29] whereas other studies ascribed the enhanced activity to a synergistic electronic or catalytic interaction of the individual elements.[30] An efficient utilization of catalysts is generally obtained by tuning the morphology of the alloy, *e.g.* by the formation of nanoporous materials with large specific surface area and thus a higher number of active sites. In literature, galvanic replacement reactions or dealloying processes were used to produce porous AuPt alloys.[31–33] Namely, dealloying procedures that selectively dissolve one metal component in the alloy were reported for AuAgPt that led to a bulk nanoporous morphology.[34–37] In these studies, the main goal was to produce nanoporous Au with the intent that Pt addition would prevent the typical coarsening of porous Au; this due to Pt blocking rapid Au surface diffusion during dealloying. In contrast to these bulk approaches, to the best of our knowledge there are no reports up to now on producing nanoporous AuPt alloys as nanoparticles by a dealloying process.

In the present work, we present a combination of optimized dewetting of a triple element, Au, Pt, Ag-layer configuration on an ordered $TiO_2$ substrate to form alloyed NPs. This is then followed by selective dissolution of Ag that allows for the formation of homogeneous porous alloyed AuPt NPs. Dewetting and dealloying form these nanoporous AuPt alloy NPs directly on regularly ordered TiO$_2$ NTs. In a first step, alloying of Au, Pt and Ag is achieved through thermal dewetting of sputtered metal layers (1 nm-thick Au, 1 nm-thick Pt and 6 nm-thick Ag) as illustrated in Scheme 1. Then dealloying was carried out by selective chemical dissolution of Ag in HNO$_3$, leading to porous AuPt alloy nanoparticles with a diameter in the 10-70 nm range, decorated on the TiO$_2$ nanotube layers. These porous alloy-decorated TiO$_2$ tubes were investigated for photocatalytic H$_2$ generation where the nanoporous co-catalyst provided a significant enhancement in comparison with either single metal or the non-porous alloy as will be shown below.

**Figure 1a-c** shows top and cross-sectional view scanning electron microscopy (SEM) images of the spaced TiO$_2$ NTs used in this work. Layers of these TiO$_2$ NTs were formed by an electrochemical anodization of a Ti sheet in a triethylene glycol electrolyte (see experimental section). The NTs have an average diameter of 200 nm and a length of 3 µm. An important feature of this type of ordered NT-layers is that the individual NTs are grown with an interspace of ≈150 nm between each other.[38] This spacing allows for a defined decoration of noble metal particles on the tube mouth and walls and guarantees the access of light and electrolyte into the interspace. In order to form a noble metal decoration, the TiO$_2$ NT samples were coated with Au, Pt and Ag (in the sequence of 1 nm Au, 1 nm Pt and 6 nm Ag, denoted as Au$_1$Pt$_1$Ag$_6$, Figure 1d) by a plasma sputtering system (Leica EM SCD500). This sequence was found to be most effective in obtaining uniform dewetting. Moreover, the metal layers were specifically coated on the outer walls of TiO$_2$ NTs because sputtering was carried out at a shallow angle.[13,39] Clearly, the metal layers homogeneously cover the surface of TiO$_2$ NTs which is a prerequisite for the success of the subsequent thermal dewetting process.

For dewetting, the AuPtAg-decorated TiO$_2$ nanotubular samples were annealed in Ar atmosphere. The treatment induces simultaneously a dewetting of the metal layers and alloying of the elements. The use of Ar gas for alloying is essential to prevent the oxidation of Ag during the thermal treatment. Among the three metals, Pt has the highest melting point, therefore we carried out the dewetting process of the sputtered metal films at 500 °C which was found to be sufficient to trigger dewetting of Pt.[14] Figure 1e shows an SEM image of the formed AuPtAg alloy decoration on the TiO$_2$ NTs.

These AuPtAg alloy decorated samples were then used for a chemical treatment in HNO$_3$. This induces dealloying by a selective dissolution of Ag and a porous structure is formed that is enriched in the more noble elements, Au and Pt.[40] Figure 1f shows an SEM image of the particles after dealloying. Clearly, the nanoparticles become porous and have a smaller diameter than the particles before alloying.

**Figure 2a** shows a TEM image of a TiO$_2$ nanotube after sputtering the layered Au$_1$Pt$_1$Ag$_6$. It is again apparent that the metal layers uniformly coat the TiO$_2$ tube wall although there are some openings in the layers due to the low thickness of the sputtered layers (Figure S1). After thermal dewetting in Ar, AuPtAg alloyed NPs of various sizes form on the nanotube walls (Figure 2b). The homogeneity of these particles was confirmed by TEM-EDS mappings as shown in Figure 2f1-f4. Obviously, Au, Pt and Ag are evenly distributed within the nanoparticle (in case of Figure 2f1, the particle has a diameter of approximately 70 nm). Not only one particle but all the investigated NPs (shown in Figure S2) consist of Au, Pt and Ag, suggesting that thermal dewetting is a feasible approach to produce fully alloyed AuPtAg NPs at a much lower temperature than the melting points of metal films (T$_{m_{Pt}}$ = 1769 °C).[41] Figure 2c and 2g show a TEM image and EDS mappings of porous alloy NPs on an TiO$_2$ nanotube. A typical pore size in the range of 2-5 nm is formed which demonstrates successful dissolution from the AuPtAg alloys without significant coarsening. Moreover, EDS mappings indicate the loss of Ag from the alloys after the HNO$_3$ treatment (Figure 2g1-g4 and S3). HRTEM images in Figure 2d,e further demonstrate the formation of AuPtAg alloyed particles. The lattice spacing of the (111) plane is 2.34 Å which lies between the values for pure Au and Ag (2.35 Å) and Pt (2.28 Å).[42] The observed intermediate spacing supports the formation of a truly alloyed structure. For the uncoated part of the tube, a lattice spacing of 3.53 Å can be found which can be assigned to anatase (101).[43]

**Figure 3a** and **b** show XRD patterns of TiO$_2$ samples after every step. Prior to metal deposition, the as-anodized TiO$_2$ nanotubular samples were annealed at 450 °C for 1 h in air to crystallize the structures. The XRD patterns reveal mainly anatase phase (with an intense (101) peak at 2θ = 25.3°) and a small portion of rutile phase (a weak (110) peak at 2θ = 27.4°). The successful deposition of noble metals (5 nm-thick Au, 5 nm-thick Pt and 30 nm-thick Ag) was confirmed by the appearance of a broad (200) peak at 2θ = 44.3° (green curve, JCPDS no. 00-001-1167) which is the reflection of mainly Ag (Au and Ag have the same unit cell and similar lattice constants, therefore they have a very similar XRD reflection).[44] After thermal dewetting, the XRD pattern (red curve) shows an increase in the intensity of the metal peaks. Moreover, the peak is shifted to 2θ = 44.5° which lies between the Au(200)-Ag(200) and Pt(200) peak (2θ = 45.6°, JCPDS no. 01-088-2343),[45] indicating the successful formation of AuPtAg alloyed NPs. The XRD pattern of dealloyed sample exhibits a significant decrease in the intensity of the metal peak at 2θ = 44.5° which is due to the dissolution of Ag during the chemical treatment in nitric acid. This is confirmed by EDS measurements (Table S1) that show a substantial reduction in the Ag concentration to 0.14 at% (after dealloying), compared to 1.64 at% before dealloying. The small amount of Ag that remains in the nanoporous alloy particles is ascribed to the presence of Pt in the alloys that reduces Au mobility and thus decreases a further progress of Ag depletion.[34–37] The amounts of Au and Pt slightly decrease due to some loss by rearrangement of atoms during dealloying process.[40]

XPS measurements were carried out to determine the chemical states of the constituent atoms in the samples. Figure 3c-e show XPS spectra of Au4f, Pt4f and Ag3d measured for TiO$_2$ NTs decorated with Au$_1$Pt$_1$Ag$_6$ after individual steps (sputtering, dewetting and dealloying). All metal peaks were shifted to lower binding energies after dewetting and dealloying, indicating the formation of alloyed NPs.[30] Specifically, after deposition the binding energies of Au4f$_{5/2}$ and Au4f$_{7/2}$ are 88.2 and 84.4 eV respectively, *i.e.* well in line with elemental Au,[46] whereas after dealloying these peaks shift to 87.5 and 83.9 eV (Figure 3c). In analogy, the Pt peaks (Pt4f$_{5/2}$ and Pt4f$_{7/2}$), centered at 74.7 and 71.4 eV after deposition,[47] are after dealloying shifted to 74.2 and 71.0 eV. Also the binding energy (BE) difference between Au and Pt peaks after dealloying became slightly smaller (13.5 eV and 13.3 eV, respectively). These peak shifts are usually reported for alloy formation.[48] It is also noted that the intensity of Ag peaks measured for dealloyed samples remarkably decreases as expected due to selective etching.

The TiO$_2$ NTs decorated with the alloy or porous alloy NPs were then investigated for photocatalytic H$_2$ generation from the aqueous-ethanol solutions under UV light irradiation ($\lambda$ = 365 nm, see experimental section). **Figure 4a** shows the H$_2$ evolution rate of TiO$_2$ NTs and TiO$_2$ NTs decorated with pure Ag, pure Au, pure Pt, AuPtAg alloy and porous alloy particles. Pure TiO$_2$ NTs deliver a very low H$_2$ rate of 0.3 µL h$^{-1}$ owing to a fast electron-hole recombination. Decorating the NTs with Ag NPs slightly increases the efficiency (1.4 µL h$^{-1}$), while Au and Pt nanoparticle decoration (Figure S4) significantly enhance the photocatalytic H$_2$ generation more than 100 times (27.4 and 35.9 µL h$^{-1}$, respectively). This may, in part, be ascribed to the formation of a Schottky junction between the noble metal particles and TiO$_2$, which leads to an efficient electron-hole separation. The better performance of Au and Pt in photocatalysis is often assigned to their higher work function compared to Ag ($\Phi_{Pt}$ = 5.93 eV, $\Phi_{Au}$ = 5.31 eV and $\Phi_{Ag}$ = 4.74 eV).[49] Furthermore, in line with previous reports,[13,14,50] a thermal dewetting to convert noble metal layers (Ag, Au and Pt) into NPs enhances the H$_2$ production efficiency due to a higher open TiO$_2$ area that is needed to allow spatially separated hole exit to the electrolyte.[19,51,52] Interestingly, using three metal decoration (sputtered Au$_1$Pt$_1$Ag$_1$) or alloy nanoparticles after dewetting (Au$_1$Pt$_1$Ag$_1$ alloy, Figure S5) show only a small increase in photocatalytic H$_2$ evolution compared to the elements, whereas porous Au$_1$Pt$_1$Ag$_1$ alloy NPs (after dealloying) remarkably enhance the $H_2$ generation up to 82.0 µL h$^{-1}$ (this is a doubling compared to the non-dealloyed AuPtAg NPs).

To gain more insight into the effect of Ag on the porosification of alloyed NPs and photocatalysis, $TiO_2$ NTs were decorated with the same amount of Au and Pt (1 nm-thick) and various amounts of Ag. All these samples were thermally dewetted in Ar and dealloyed in nitric acid. The results in Figure 4b show that increasing the thickness of the sputtered Ag films also enhances the evolved $H_2$ gas. For instance, $Au_1Pt_1Ag_6$ delivers a $H_2$ rate of 138.5 µL h$^{-1}$, that is an almost 70% enhancement compared to samples decorated with $Au_1Pt_1Ag_1$ (82.0 µL h$^{-1}$). In line with common dewetting mechanisms,[53] the size of porous NPs grows with an increasing Ag layer thickness. In our experiments, $Au_1Pt_1Ag_6$ was found to be an optimized condition for photocatalytic $H_2$ evolution (Figure S6). Larger amounts of Ag ($Au_1Pt_1Ag_{12}$) lead to larger porous NPs which is detrimental for the $H_2$ evolution.[14] Furthermore, when the ratio between Au/Pt and Ag is fixed (1:1:6), the total amount of sputtered metal layers was found to have a considerable influence on the $H_2$ efficiency (Figure 4c). In our experiments, clearly the 1 nm Au, 1 nm Pt and 6 nm Ag represent an optimized condition for a maximized $H_2$ evolution rate – larger amounts of loaded noble metals result in a larger nanoparticle size which increasingly prevents efficient light absorption in the $TiO_2$ (Figure S7). Porous AuPt alloy co-catalysts also outperform the $H_2$ generation from similar amounts of pure Au and Pt co-catalysts ($Au_2$ and $Pt_2$, Figure S8). Please note that the sputtering conditions for monometallic decoration used in this work (Au and Pt layers of a nominal thickness of 1-2 nm) have been optimized for maximized photocatalytic $H_2$ generation in previous work.[13,14,54,55] In comparison with literature, the photocatalytic $H_2$ generation in the current work shows an enhanced performance (when compared under similar experimental conditions – table S2).

As anticipated, the temperature of thermal dewetting was found to be very important to generate an active catalyst. Figure 4d shows the amounts of $H_2$ gas produced by $TiO_2$ NTs that were loaded with $Au_1Pt_1Ag_6$ and dewetted at different temperature in Ar before being dealloying in nitric acid. Annealing at 400 °C is not sufficient to induce dewetting, therefore only non-porous alloy particles were observed (Figure S9a). Thermal treatment at higher temperatures can also provide dewetting and allow the formation of porous NPs (Figure S9b-d). However, XRD patterns in Figure S10 shows that annealing at 550 °C causes larger amounts of rutile phase being formed in the $TiO_2$ substrate, and this in turn leads to a decrease in the $H_2$ generation rate. The decreased efficiency is attributed to porosity loss (sintering phenomena) and by conversion of anatase to rutile (being less active polymorph).[14,56–58] The highest photocatalytic $H_2$ efficiency was obtained for thermal dewetting at 500 °C.

In comparison with mono-metallic decoration (Ag, Au and Pt), $TiO_2$ NTs decorated with porous AuPt alloy NPs yield a much higher efficiency (97, 5 and 4 times, respectively). This can be mainly attributed to two factors: (*i*) the synergistic effects between Au and Pt in the alloys, and (*ii*) the increased specific surface area of the co-catalyst particles due to the formation of nanopores. Commonly the beneficial effect of an alloyed AuPt catalyst are ascribed to the difference in the electronegativity between Au and Pt that influences the electron density by adjusting the Fermi level of the co-catalyst junction to be more favorable for charge separation and an enhanced photocatalytic efficiency.[30] Another key factor is that Pt significantly enhances the hydrogen atom adsorption and recombination, however Pt also catalyzes the reverse reaction. As a result, Pt significantly reduces the overpotential for hydrogen evolution.[59–61] In contrast, Au shows no strong interaction with $H_2$ and therefore the presence of Au atoms can play an important role in the desorption of generated $H_2$ on the alloys.[62] Our results however show that even more important is the porosity *i.e.* the porous alloyed NPs with a pore size less than 5 nm provides a significantly improved active surface area and therefore leads to a strongly improved $H_2$ production.[34] For Au-containing co-catalysts in literature, beneficial effects were often related to plasmon resonance.[28] However, the high photocatalytic efficiency of nanoporous alloy-decorated $TiO_2$ NTs in the present work cannot be ascribed to plasmon-related effect. Specifically, reflectance results (Figure S11a) measured for TiO$_2$ NTs and NTs decorated with noble metals show that porous AuPt alloy NP-decorated TiO$_2$ NTs show a higher optical absorption between 400 and 500 nm, this cannot be a significant contribution as in our case the photocatalytic measurements were conducted by irradiating samples with UV light (λ = 365 nm) – at this wavelength a similar absoprtion is obtained for all samples. Together with photocurrent measurements at different wavelengths (Figure S12), this indicates that apart from an enlarged co-catalyst surface area, only electronic and chemical effects of AuPt are essential for the effects observed from this work.

In summary, we show that porous AuPt alloy NPs can be formed on spaced TiO$_2$ NTs by a simple method that combines thermal dewetting and dealloying. For this, three thin metal layers (Au, Pt and Ag) were sputter-coated on TiO$_2$ NTs, then thermally dewetted which formed homogeneous AuPtAg alloyed NPs. These particles were then porosified by a dealloying process using selective chemical dissolution of Ag. The porous AuPt alloyed NP-decorated TiO$_2$ NTs exhibit a significantly higher photocatalytic H$_2$ evolution than mono-metallic-decorated TiO$_2$ NTs under UV light irradiation. The approach demonstrated here not only provides a promising step to produce alloyed NPs (at much lower temperatures compared to the melting point) but it represents a basic strategy to create locally defined, supported porous alloyed NPs for catalytic or co-catalytic applications.

**Acknowledgements**

We would like to acknowledge the ERC, the DFG, the Erlangen DFG cluster of excellence EAM, project EXC 315 (Bridge), the DFG funCOS and the Operational Programme Research, Development and Education-European Regional Development Fund, project no. CZ.02.1.01/0.0/0.0/15_003/0000416 of the Ministry of Education, Youth and Sports of the Czech Republic for financial support.

# Table of contents

Ordered TiO$_2$ nanotubes decorated with nanoporous AuPt alloyed nanoparticles show a significant enhancement of photocatalytic H$_2$ generation in comparison with either the single metal or non-porous alloy.

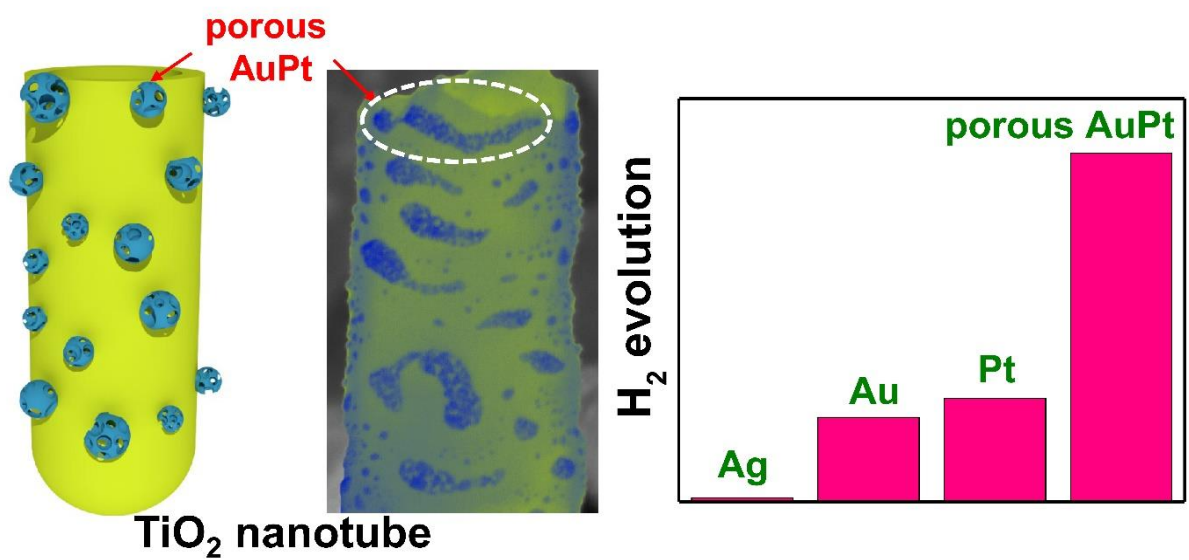

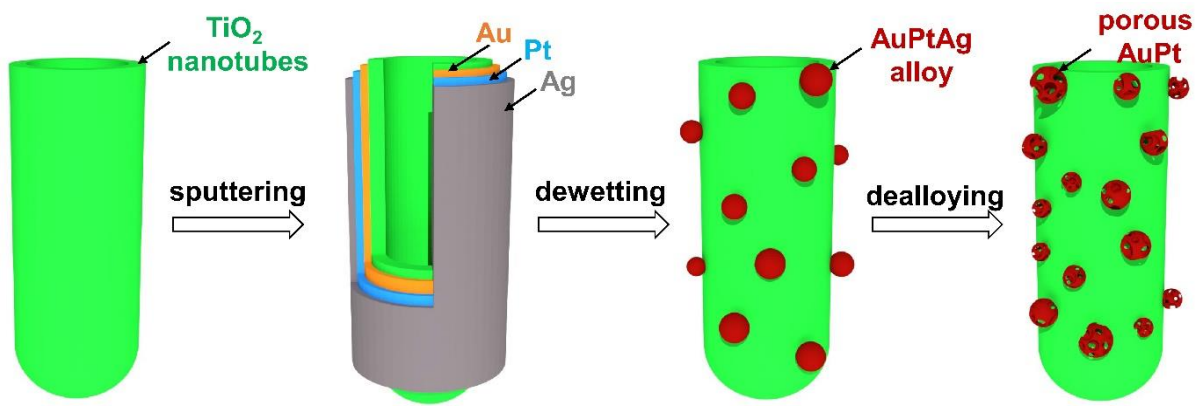

**Scheme 1.** Formation of porous AuPt alloyed NPs on TiO$_2$ NTs. Noble metal was deposited on TiO$_2$ NTs by a plasma sputtering machine. Dewetting was carried out by a thermal treatment in Ar and dealloying was conducted by selective chemical dissolution in nitric acid.

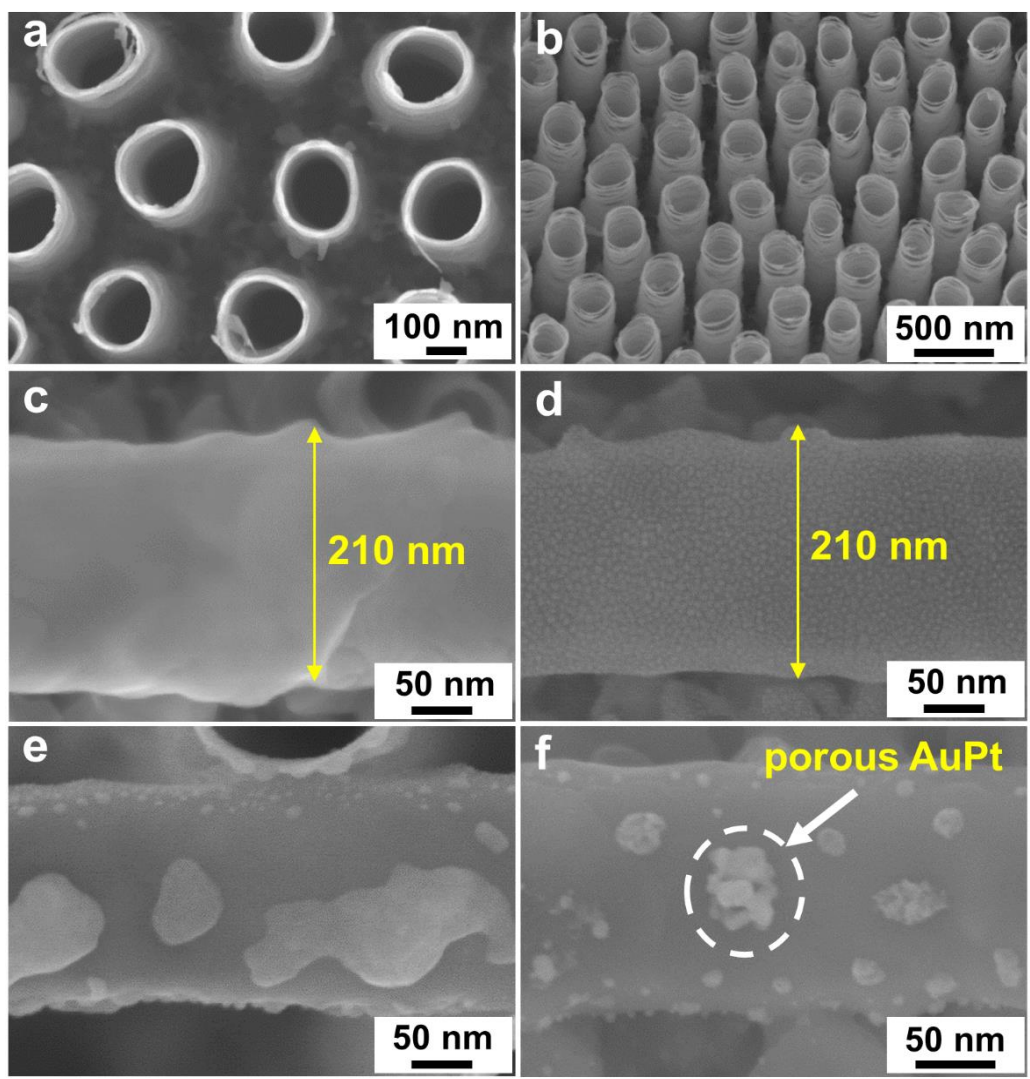

**Figure 1.** SEM images of: (a,b) spaced $TiO_2$ nanotubes, (c) cross-sectional nanotube, (d) nanotube after sputtering of 1 nm Au – 1nm Pt – 6 nm Ag ($Au_1Pt_1Ag_6$), (e) nanotube after sputtering-dewetting of $Au_1Pt_1Ag_6$ and (f) nanotube after sputtering-dewetting-dealloying of $Au_1Pt_1Ag_6$. Dewetting was conducted in Ar at 500 °C for 1 h and dealloying was carried out in $HNO_3$ at 25 °C for 4 h.

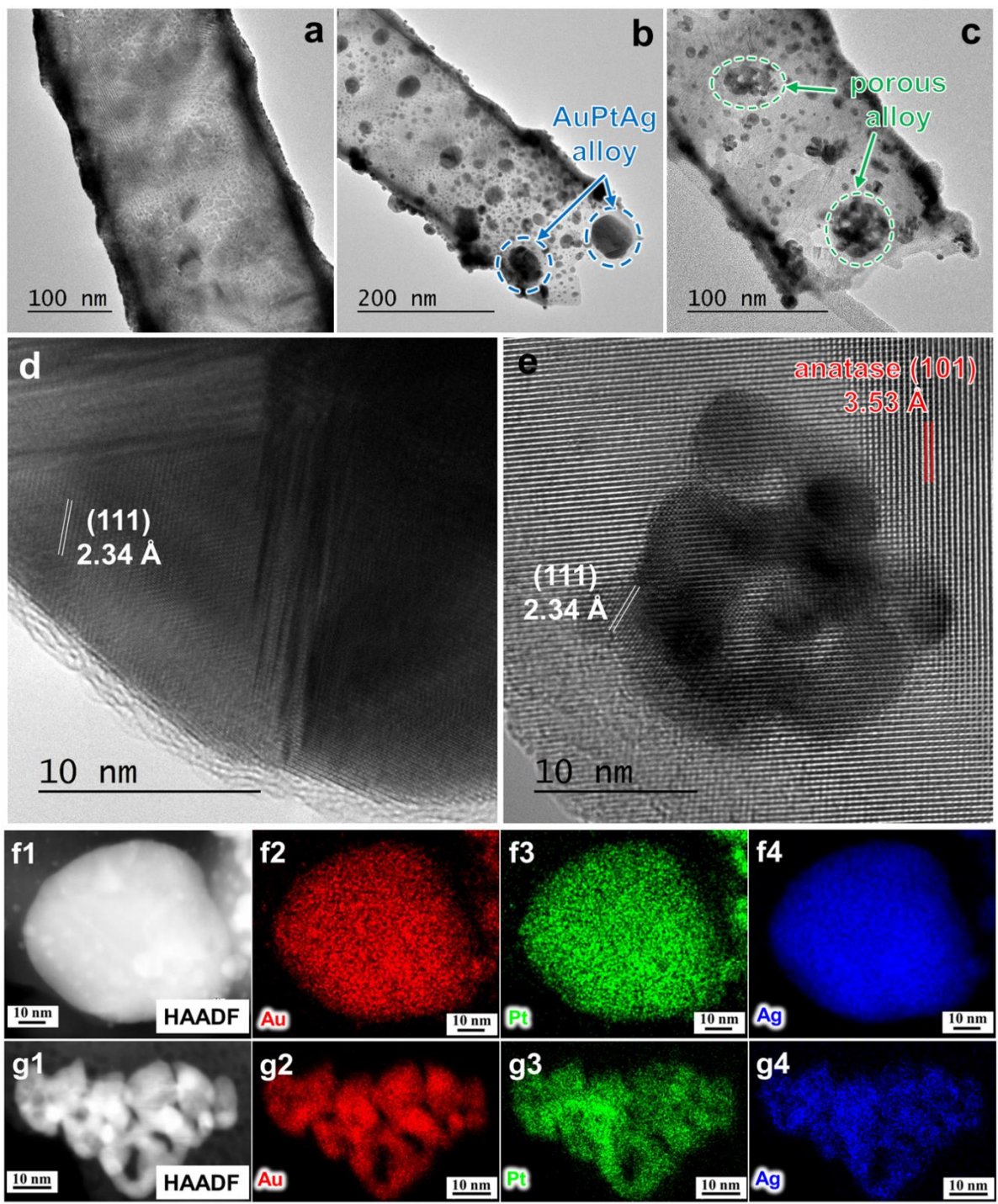

**Figure 2**. TEM images of TiO$_2$ nanotube after: (a) sputtering 1 nm-thick Au, 1 nm-thick Pt and 6 nm-thick Ag; (b) dewetting in Ar at 500 °C for 1 h; (c) dealloying in HNO$_3$ at 25 °C for 4 h; (d) AuPtAg alloy and (e) porous alloy particle. HAADF images and elemental mappings of (f1-f4) AuPtAg alloy particles and (g1-g4) porous alloy particles.

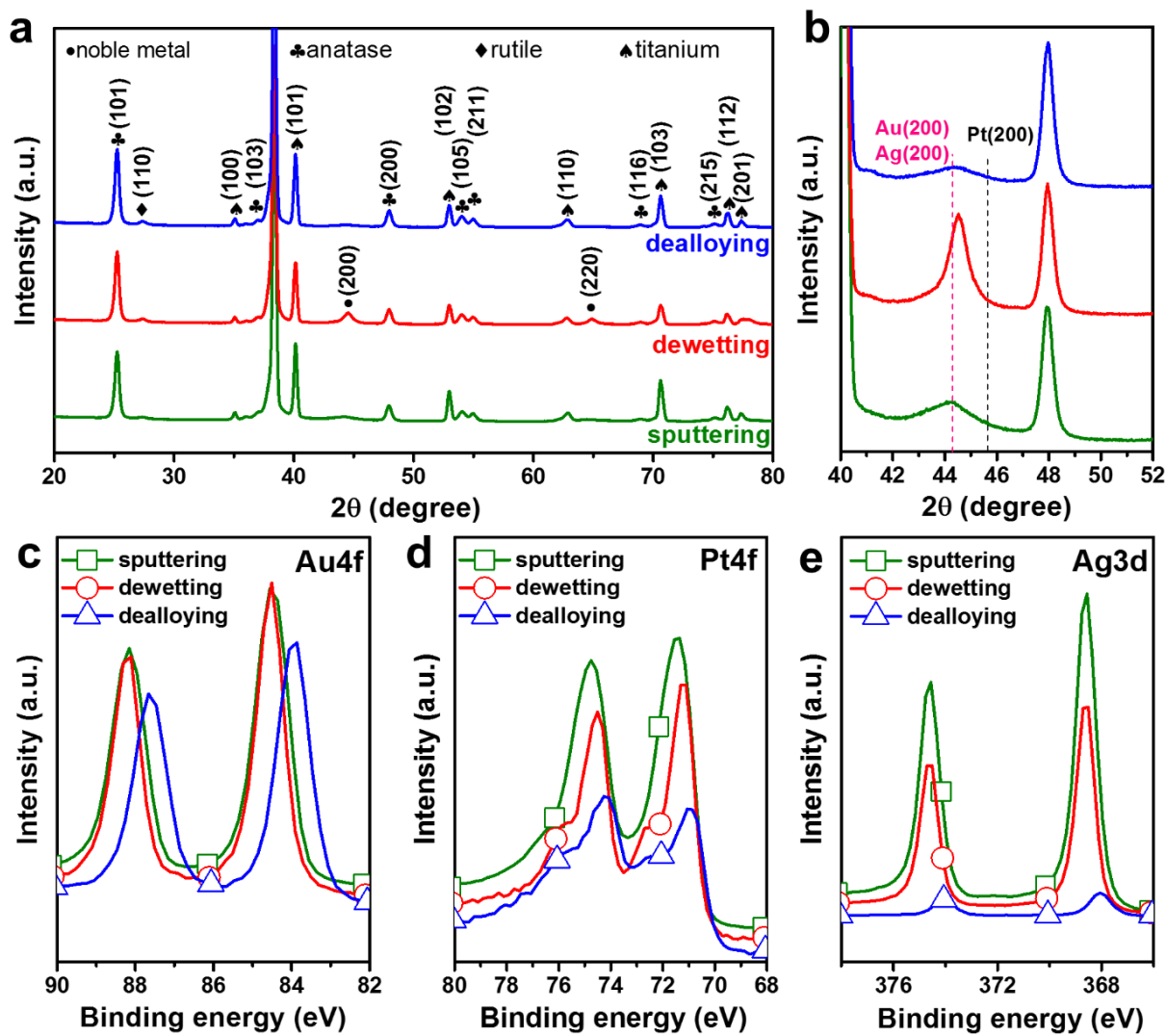

**Figure 3.** (a,b) XRD patterns of TiO$_2$ NTs; TiO$_2$ NTs after metal sputtering (5 nm-thick Au, 5 nm-thick Pt and 30 nm-thick Ag); TiO$_2$ NTs after metal sputtering-dewetting (in Ar at 500 °C for 1 h); TiO$_2$ NTs after metal sputtering-dewetting-dealloying (in HNO$_3$ at 25 °C for 4 h). (c,d,e) XPS high resolution spectra of Au4f, Pt4f and Ag3d measured for TiO$_2$ NTs after sputtering, sputtering-dewetting, sputtering-dewetting-dealloying (1 nm-thick Au, 1 nm-thick Pt and 6 nm-thick Ag).

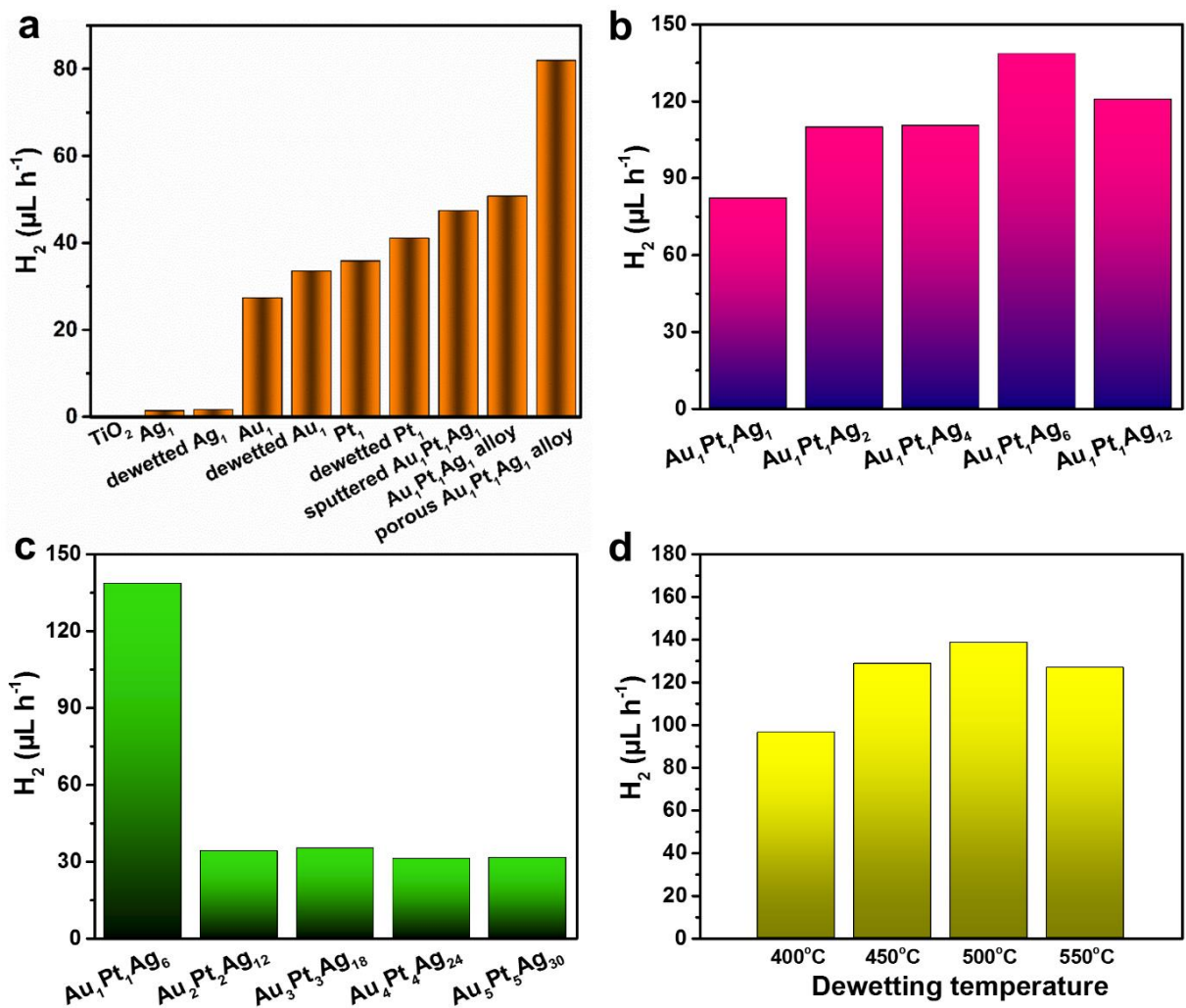

**Figure 4.** Photocatalytic H$_2$ generation of: (a) TiO$_2$ NTs and TiO$_2$ NTs decorated with 1 nm-thick Ag / 1 nm-thick Au / 1 nm-thick Pt (dewetting in Ar at 500 °C for 1 h; dealloying in HNO$_3$ at 25 °C for 4 h); (b) TiO$_2$ NTs decorated with 1 nm-thick Au, 1 nm-thick Pt and different amounts (1, 2, 4, 6, and 12 nm-thick) of Ag (all samples were dewetted in Ar at 500 °C for 1 h and then dealloyed in HNO$_3$ at 25 °C for 4 h); (c) TiO$_2$ NTs decorated with different amounts of Au, Pt and Ag (all samples were dewetted in Ar at 500 °C for 1 h and then dealloyed in HNO$_3$ at 25 °C for 4 h); (d) TiO$_2$ NTs decorated with 1 nm-thick Au, 1 nm-thick Pt and 6 nm-thick Ag (all samples were dewetted in Ar at different temperatures for 1 h and then dealloyed in HNO$_3$ at 25 °C for 4 h).

# Nanoporous AuPt and AuPtAg alloy co-catalysts formed by dewetting-dealloying on ordered TiO$_2$ nanotube surface lead to significantly enhanced photocatalytic H$_2$ generation

Nhat Truong Nguyen,[a] Selda Ozkan,[a] Ondrej Tomanec,[b] Xuemei Zhou,[a] Radek Zboril,[b] Patrik Schmuki[a,b,c]*

[a] Department of Materials Science and Engineering WW4-LKO, University of Erlangen-Nuremberg, Martensstrasse 7, D-91058 Erlangen, Germany.

[b] Regional Centre of Advanced Technologies and Materials, Department of Physical Chemistry, Faculty of Science, Palacky University, Slechtitelu 11, 783 71 Olomouc, Czech Republic.

[c] Chemistry Department, Faculty of Sciences, King Abdulaziz University, 80203 Jeddah, Saudi Arabia.

*Corresponding author. Email: schmuki@ww.uni-erlangen.de

**Experimental section**

- *Formation of TiO$_2$ tubes*: Titanium foils (Advent Research Materials, 0.125 mm thickness and 99.6+% purity) were cleaned by sonication in acetone, ethanol and deionized water for 10 minutes, followed by being dried under a N$_2$ gas stream. The spaced TiO$_2$ nanotubes were fabricated by anodizing cleaned titanium foils in triethylene glycol (TEG) electrolyte (consisting of 0.3 M NH$_4$F and 3 M H$_2$O) at 60 V for 1 h at 60 °C. The direct current potential was supported by a VLP 2403 pro, Voltcraft power supply. After anodization, the samples were immersed in ethanol for 2 h to remove organic remnants and then dried under a N$_2$ gas stream. These samples were annealed in air at 450 °C for 1 h (Rapid Thermal Annealer – Jipelec Jetfirst 100 – a heating and cooling rate of 30 °C min$^{-1}$).

- *Noble metal decoration*: Plasma sputtering deposition (Leica EM SCD500) was used to deposit Au, Pt and Ag on TiO$_2$ samples. The amount of sputtered metals was controlled by monitoring their nominal thickness using an automated quartz crystal film. The sputtering angle is 30°.

- *Thermal dewetting*: To form alloyed nanoparticles, metal-decorated TiO$_2$ tubes were annealed in Ar at 400 °C, 450 °C, 500 °C and 550 °C for 1 h.

- *Dealloying*: Dealloying was carried out in HNO$_3$ 63% at room temperature for 4 h.

- *Characterization of the structure*: Field-emission scanning electron microscope (FE-SEM Hitachi S4800) was employed to characterize the morphology of the TiO$_2$ samples. The chemical composition of the samples was examined by X-ray photoelectron spectroscopy (XPS, PHI 5600 US). X-ray diffraction (XRD) performed with a X'pert Philips MPD (equipped with a Panalytical X'celerator detector) using graphite monochromized Cu Kα radiation (λ=1.54056 Å) was used to analyze the crystallographic properties of the samples. Transmission electron microscopy was carried out at a double-corrected FEI Titan$^3$ Themis equipped with a SuperX detector. UV-Vis diffuse reflectance spectra (DRS) were recorded on a LAMBDA 950

UV-vis spectrophotometer (Perkin Elmer, Beaconsfield, UK) with an integrating sphere ($BaSO_4$ standard white board was used as reference)

➤ *Photocatalytic measurements*: The photocatalytic $H_2$ generation experiments were carried out by irradiating with UV light (UV-LED smart Opsytec, 365 nm, 100 mW cm$^{-2}$) for 2 h. The samples were immersed in an aqueous solution of ethanol (20 vol%) in a quartz tube cell which is sealed with a rubber septum. Before photocatalytic experiments, the quartz cell was bubbled with $N_2$ gas to purge $O_2$. 200 µL gas was extracted and analyzed by gas chromatography (GCMS-QO2010SE, Shimadzu) to examine the amount of photocatalytic $H_2$ production. The GC is equipped with a thermal conductivity detector (TCD), a Restek micropacked Shin Carbon ST column (2 m × 0.53 mm). GC measurements were conducted at a temperature of 45 °C (isothermal conditions) with the temperature of the injector setup at 280 °C and that of TCD at 260°C. The flow rate of the carrier gas (Argon) was 14.3 mL min$^{-1}$.

➤ *Photoelectrochemical measurements:* IPCE spectra were recorded at a constant potential of 0.5 V (vs. Ag/AgCl) with a potentiostat (Jaissle IMP83 PC-T-BC) in $Na_2SO_4$ aqueous solutions (0.1 M) under 200-800 nm light source (Oriel 6365 150 W Xe-lamp), while the wavelength was varied using a motor driven monochromator (Oriel Cornerstone 130 1/8 m). Incident photon to energy conversion (IPCE) values were calculated using IPCE % = $(1240*i_{ph})/(\lambda*I_{light})$, where $i_{ph}$ is the photocurrent density (mA cm$^{-2}$), $\lambda$ is the incident light wavelength (nm), and $I_{light}$ is the intensity of the light source at each wavelength (mW cm$^{-2}$).

**Table S1.** Atomic concentration (evaluated from EDS results) measured for samples decorated with Au (5 nm), Pt (5 nm) and Ag (30 nm) before and after dealloying in nitric acid.

|    | before dealloying (at%) | after dealloying (at%) | Loss of noble metals before and after dealloying (%) |
|----|------|------|------|
| Ti | 64.90 | 59.63 |      |
| O  | 32.55 | 39.64 |      |
| Au | 0.44  | 0.28  | 36.4 |
| Pt | 0.48  | 0.31  | 35.4 |
| Ag | 1.64  | 0.14  | 91.5 |

**Table S2.** Comparison of our porous AuPt/TiO$_2$ NTs photocatalyst with other promising TiO$_2$ photocatalysts

| Catalyst | Amount | H$_2$ generation rate | Light source | Scavenger |
|---|---|---|---|---|
| 1 wt% PdAu/TiO$_2$ NPs[1] | 50 mg | 19.6 mol kg$^{-1}$h$^{-1}$ | 365 nm, 4×10$^7$ photons/s | 25 vol% glycerol |
| 1 at% AuPt/TiO$_2$ NTs[2] | 7 µm-long | 21.26 mol kg$^{-1}$h$^{-1}$ | 325 nm, 60 mW/cm$^2$ | 20 vol% ethanol |
| WO$_3$-Au-TiO$_2$ NTs[3] | 200 nm-long | 14.88 mol kg$^{-1}$h$^{-1}$ | 325 nm, 40 mW/cm$^2$ | 20 vol% ethanol |
| Au-C$_3$N$_4$-TiO$_2$ NPs[4] | 250 mg | 0.35 mol kg$^{-1}$h$^{-1}$ | 150 W Hg lamp | 1 vol% methanol |
| 0.15 at% Pt/TiO$_2$ NPs[5] | 7 µm-long | 18.71 mol kg$^{-1}$h$^{-1}$ | 325 nm, 60 mW/cm$^2$ | 20 vol% ethanol |
| 0.6 at% AuPt/TiO$_2$ NTs (this work) | 3 µm-long (0.225 mg)* | 27.38 mol kg$^{-1}$h$^{-1}$ | 365 nm, 100 mW/cm$^2$ | 20 vol% ethanol |

* The weight of TiO$_2$ nanotubular layers were calculated according to [6].

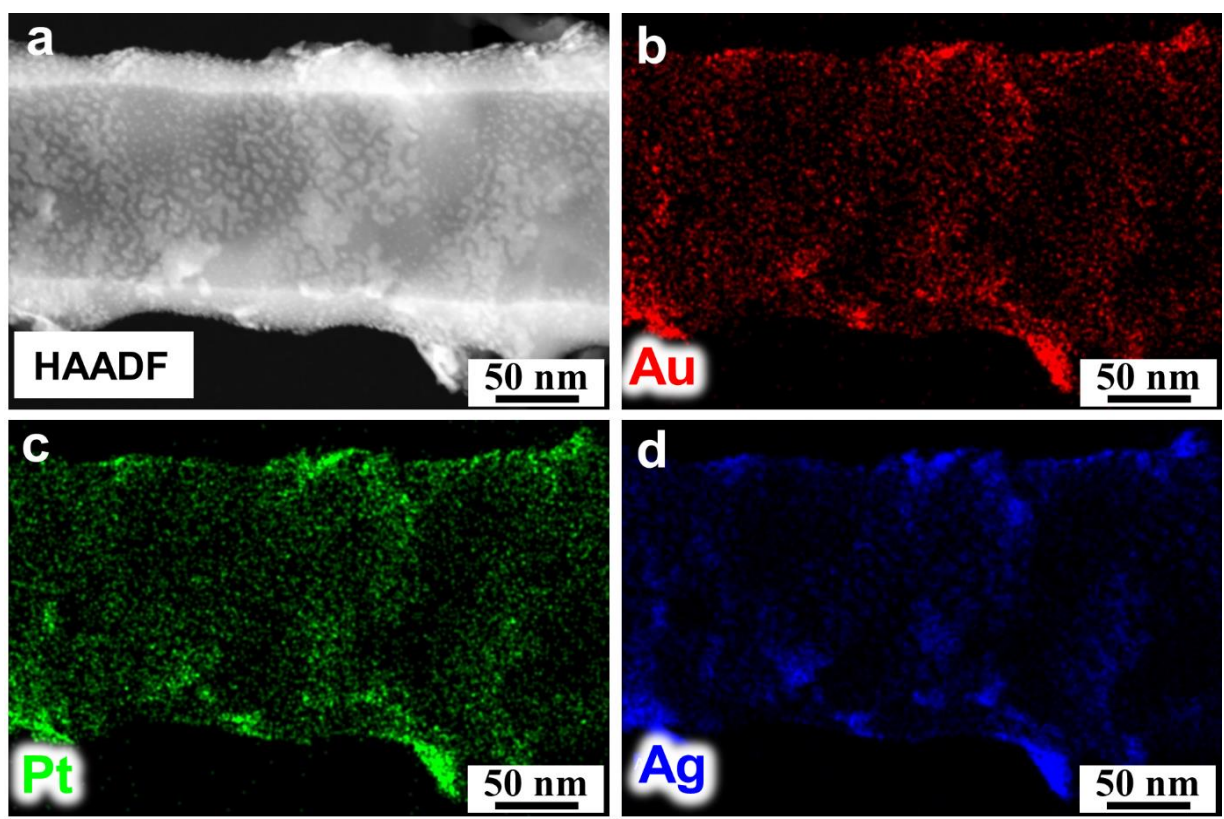

**Figure S1.** (a) HAADF image and (b-d) EDS mappings of TiO$_2$ nanotube after sputtering Au$_1$Pt$_1$Ag$_6$.

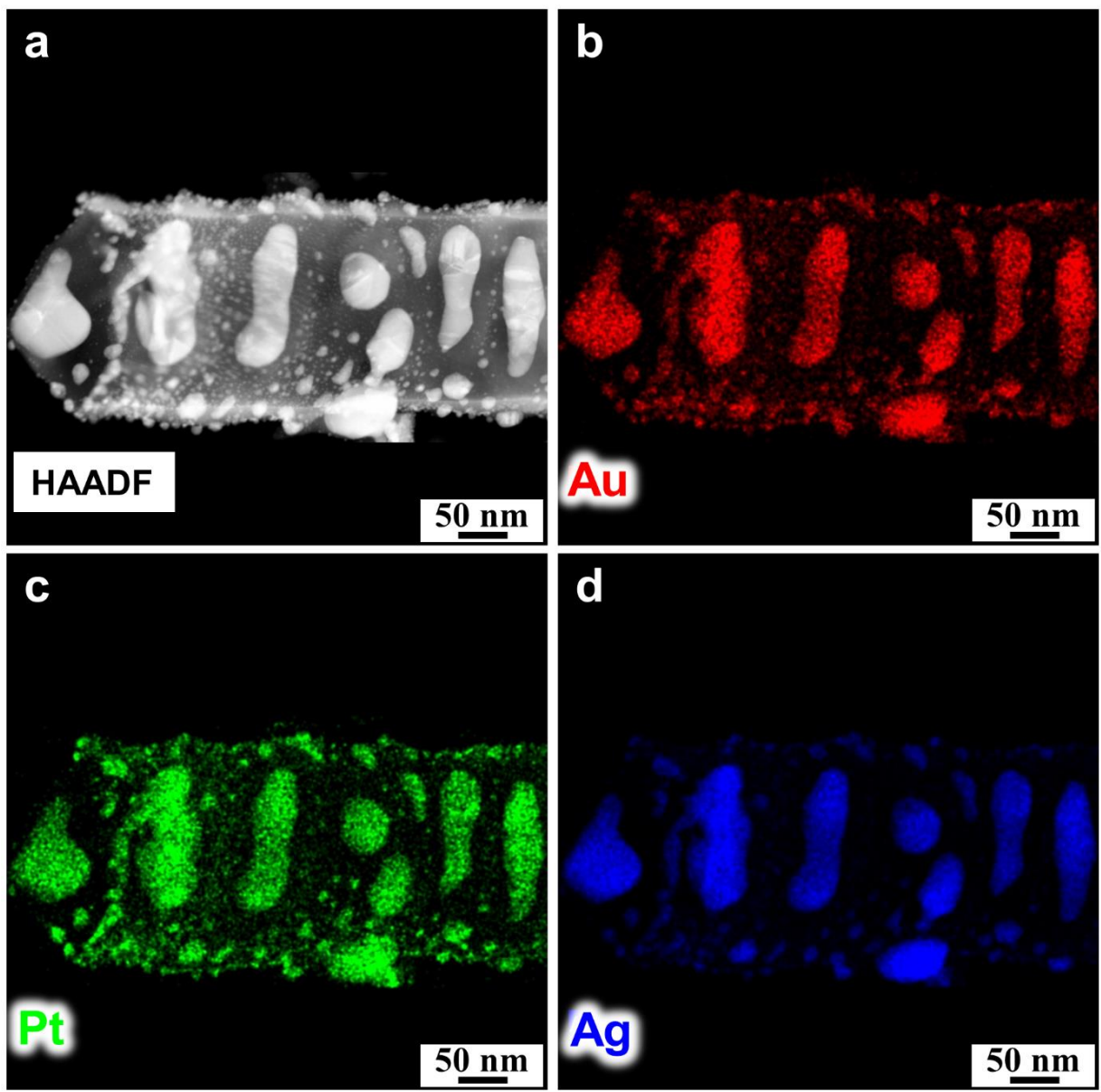

**Figure S2.** (a) HAADF image and (b-d) EDS mappings of TiO$_2$ nanotube after dewetting Au$_1$Pt$_1$Ag$_6$. Homogeneous AuPtAg alloyed nanoparticles are formed on the tube wall.

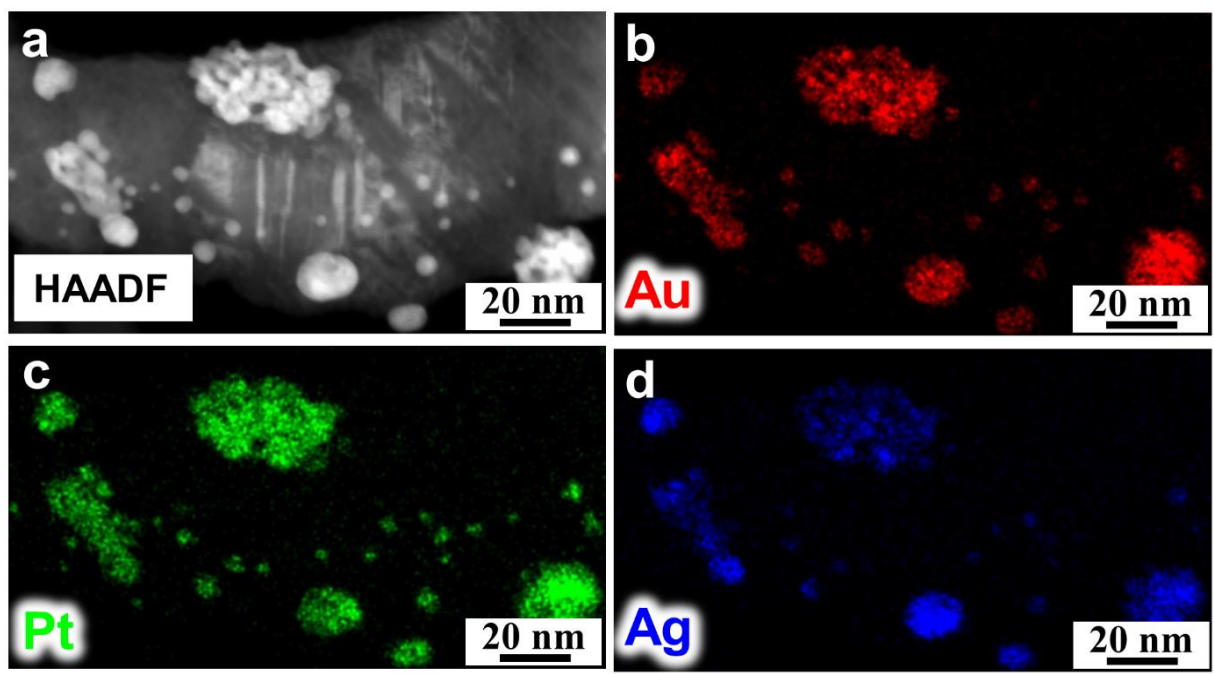

**Figure S3.** (a) HAADF image and (b-d) EDS mappings of TiO$_2$ nanotube after dealloying Au$_1$Pt$_1$Ag$_6$. Porous alloyed nanoparticles are formed on the tube wall.

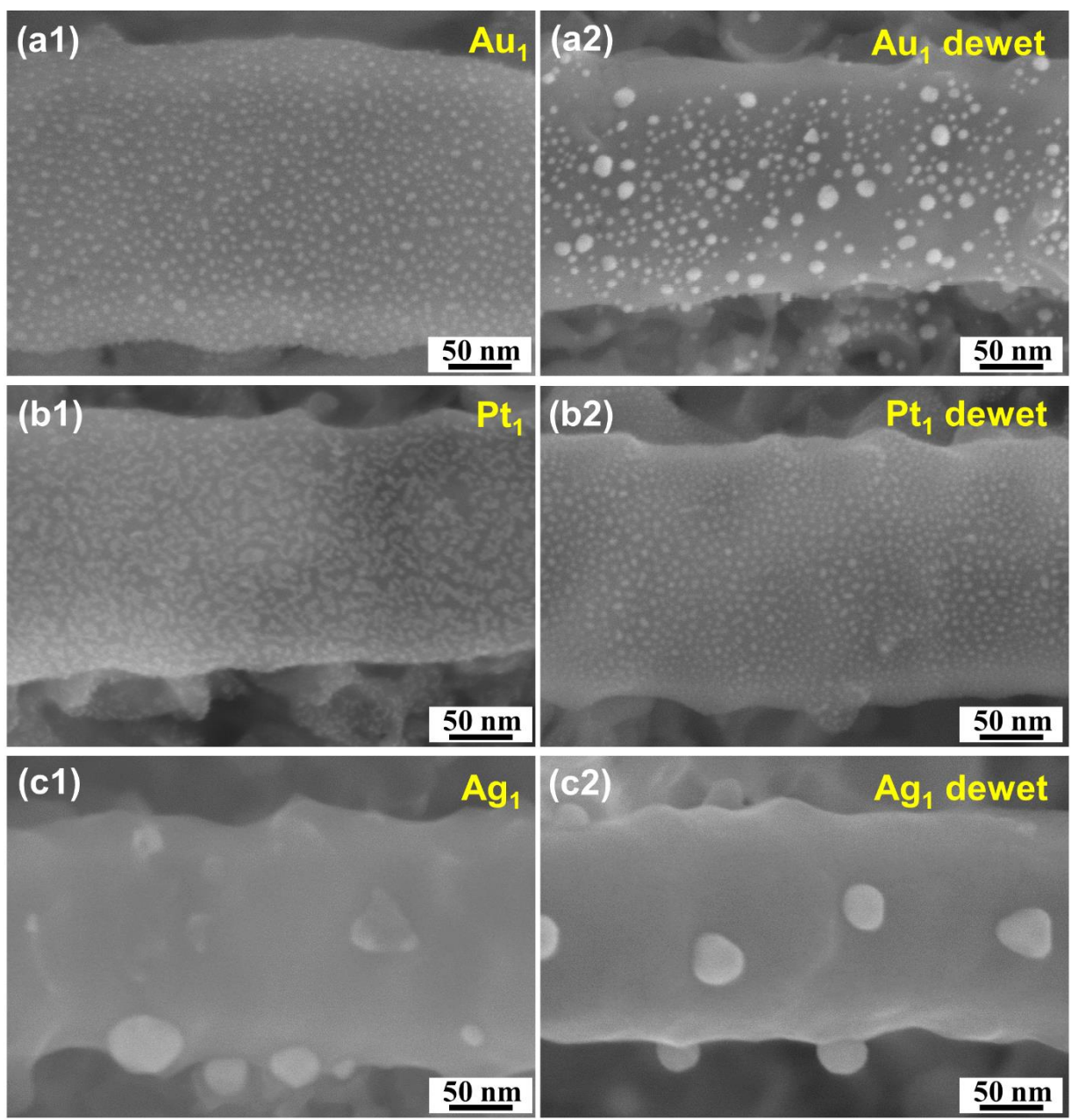

**Figure S4.** SEM images of TiO$_2$ nanotubes decorated with (a1,a2) 1 nm-thick Au; (b1,b2) 1 nm-thick Pt and (c1,c2) 1 nm-thick Ag. Samples (a2,b2,c2) were dewetted in Ar at 500 °C for 1 h.

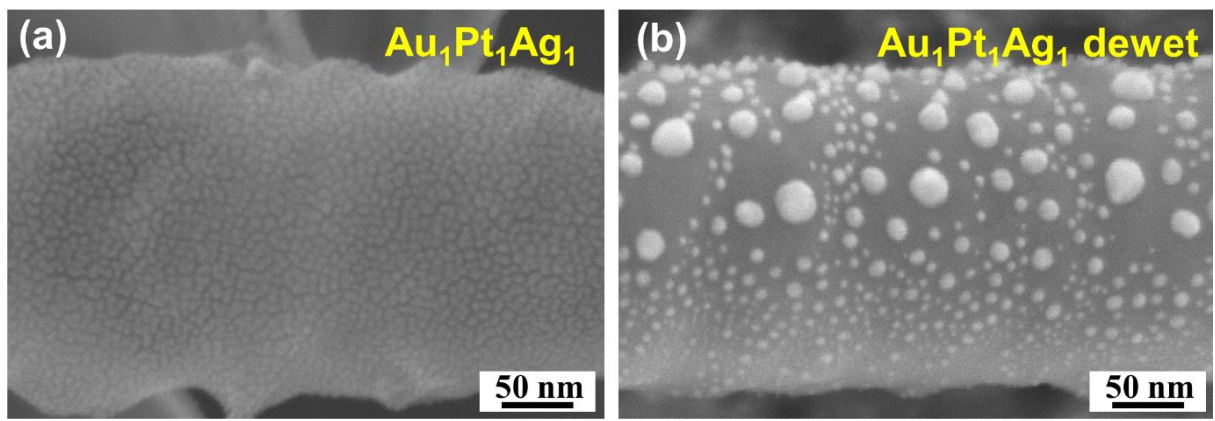

**Figure S5.** SEM images TiO$_2$ nanotubes decorated of 1 nm-thick Au, 1 nm-thick Pt and 1nm-thick Ag: (a) after sputtering and (b) after dewetting in Ar at 500 °C for 1 h.

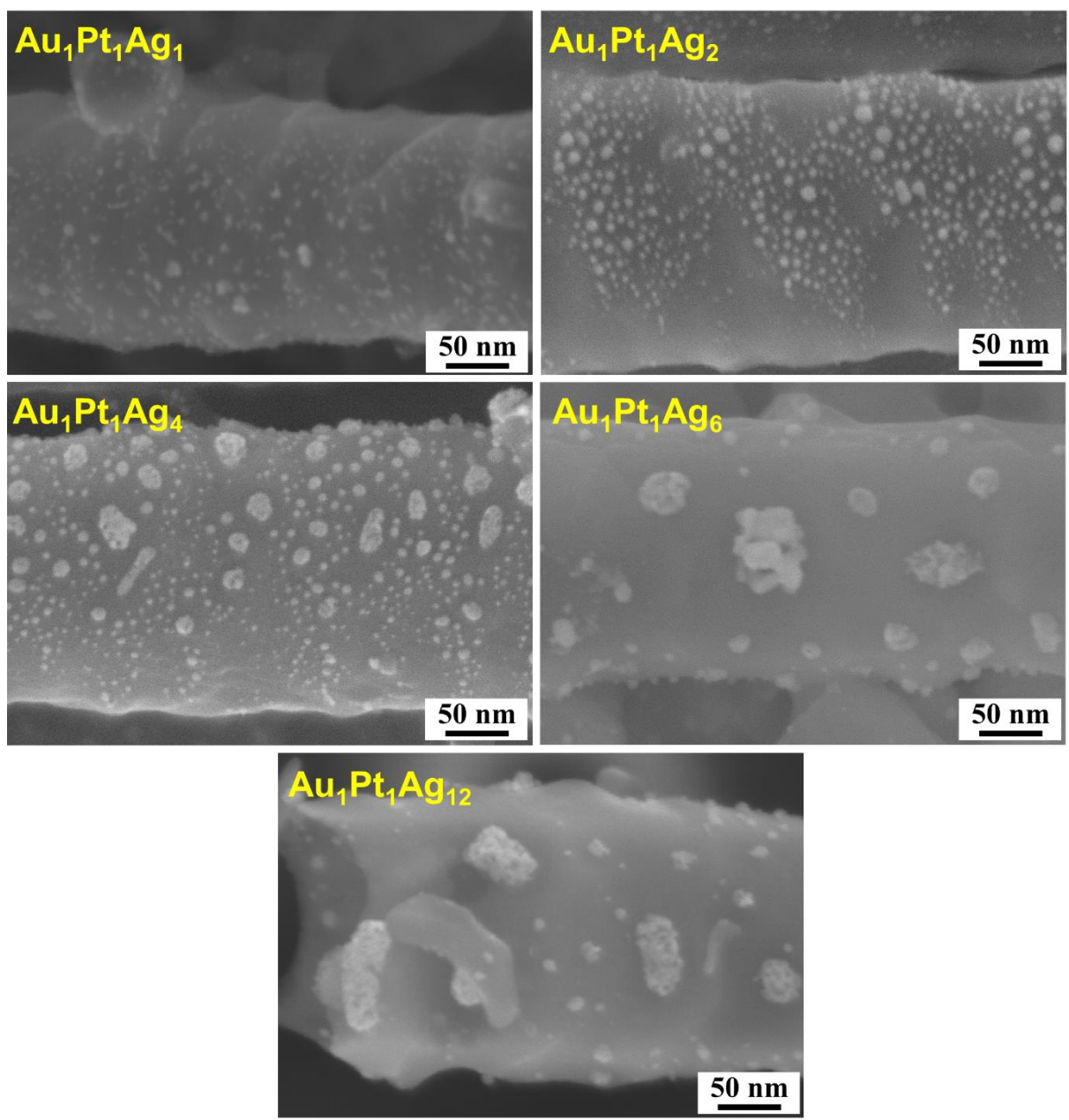

**Figure S6.** SEM images of porous AuPt alloy-nanoparticle-decorated $TiO_2$ nanotubes. All samples were sputtered with 1 nm-thick Au, 1 nm-thick Pt and different amounts of Ag (1, 2, 4, 6 and 12 nm-thick). The samples then were dewetted in Ar at 500 °C for 1 h, followed by a dealloying step in $HNO_3$ at 25 °C for 4 h.

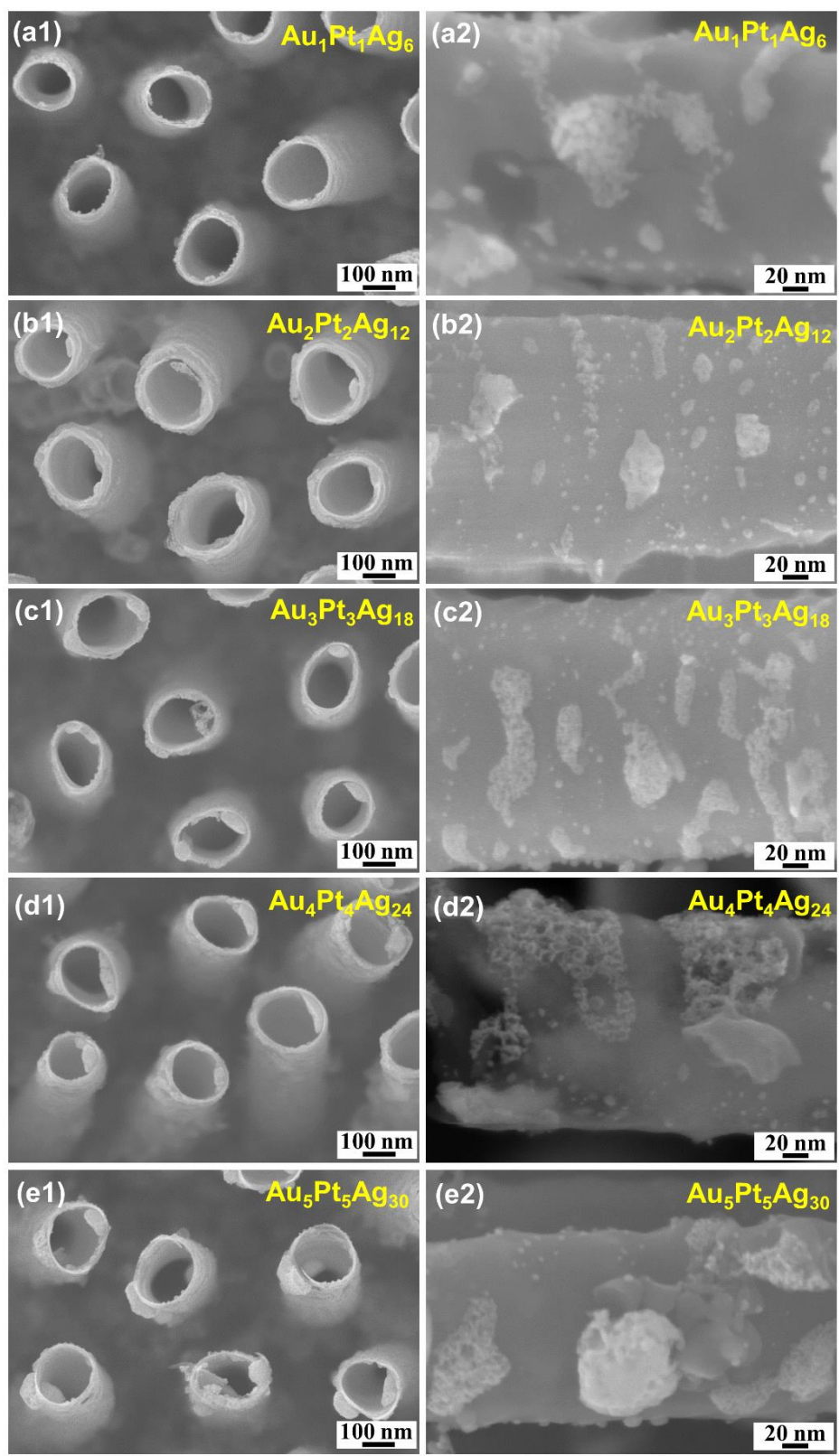

**Figure S7.** SEM images of porous AuPt alloy-nanoparticle-decorated $TiO_2$ nanotubes: (a1,a2) 1 nm-thick Au, 1 nm-thick Pt and 6 nm-thick Ag; (b1,b2) 2 nm-thick Au, 2 nm-thick Pt and 12 nm-thick Ag; (c1,c2) 3 nm-thick Au, 3 nm-thick Pt and 18 nm-thick Ag; (d1,d2) 4 nm-thick Au, 4 nm-thick Pt and 24 nm-thick Ag; (e1,e2) 5 nm-thick Au, 5 nm-thick Pt and 30 nm-thick Ag. All samples were dewetted in Ar at 500 °C for 1 h, followed by a dealloying step in $HNO_3$ at 25 °C for 4 h.

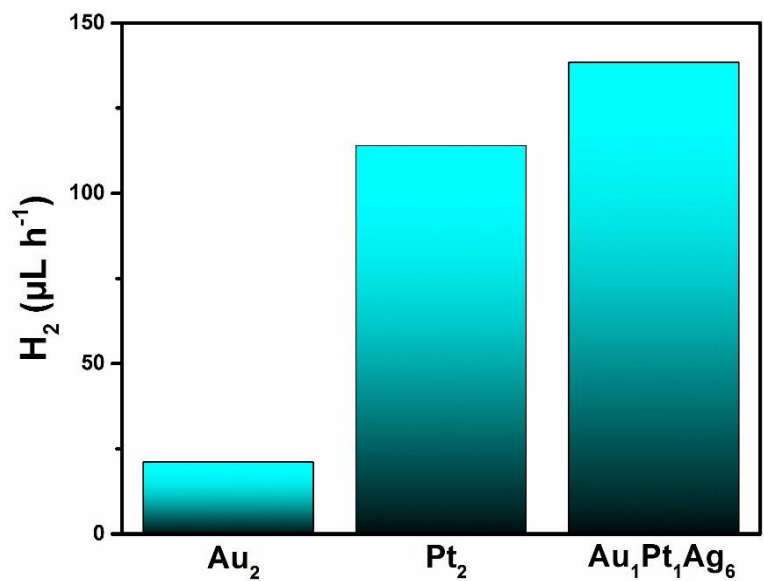

**Figure S8.** Photocatalytic $H_2$ generation of $TiO_2$ NTs decorated with 2 nm Au (Au2), 2 nm Pt (Pt2) and porous AuPt alloy particles (Au1Pt1Ag6). After sputtering, plain Au and Pt decorated $TiO_2$ tubes were annealed at 500 °C in Ar to induce dewetting.

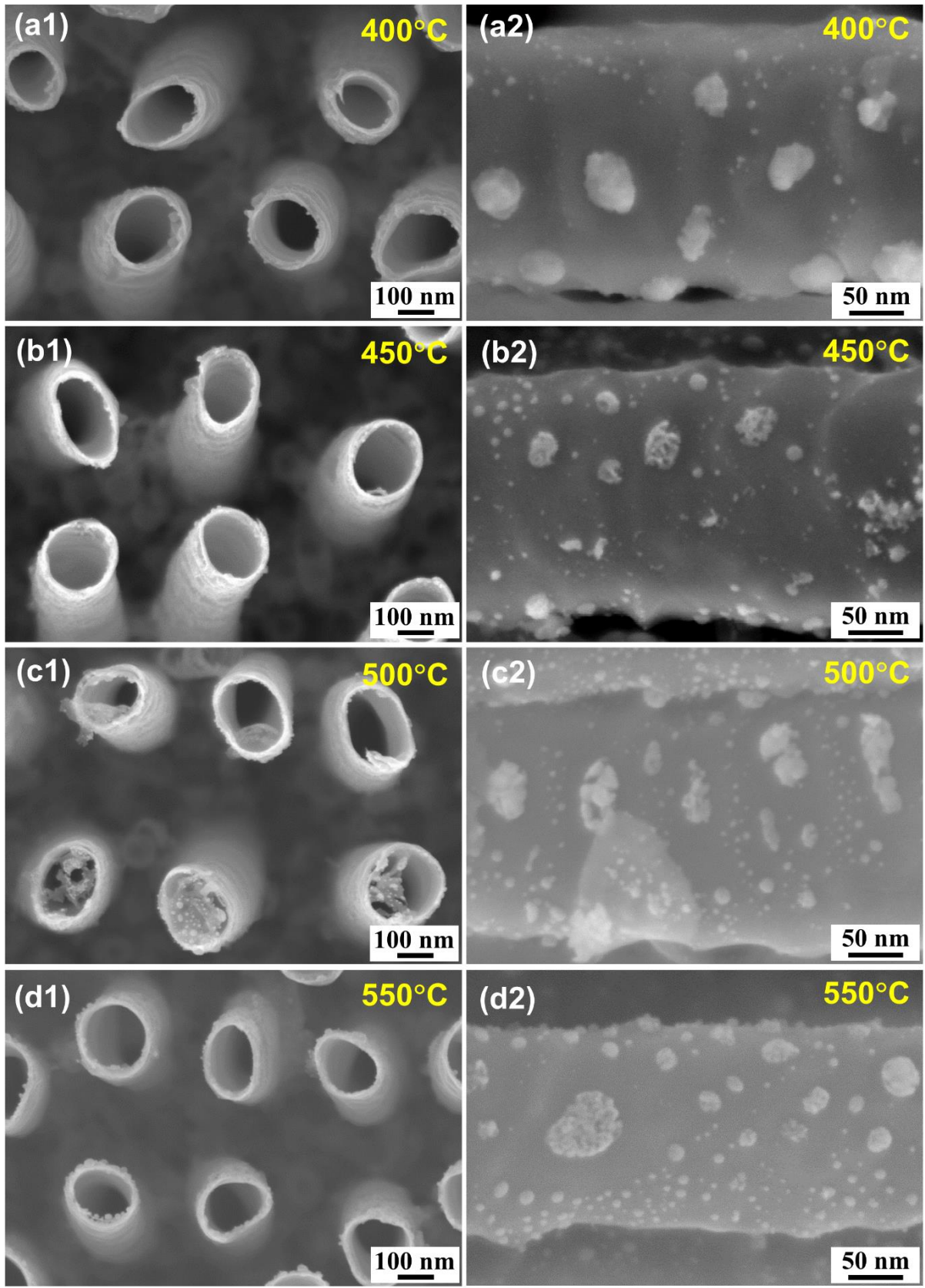

**Figure S9.** SEM images of TiO$_2$ nanotubes decorated with 1 nm-thick Au, 1 nm-thick Pt and 6 nm-thick Ag followed by a thermal dewetting in Ar for 1 h at different temperatures (a1,a2) 400 °C, (b1,b2) 450 °C, (c1,c2) 500 °C and (d1,d2) 550 °C. All samples were dealloyed in HNO$_3$ at 25 °C for 4 h.

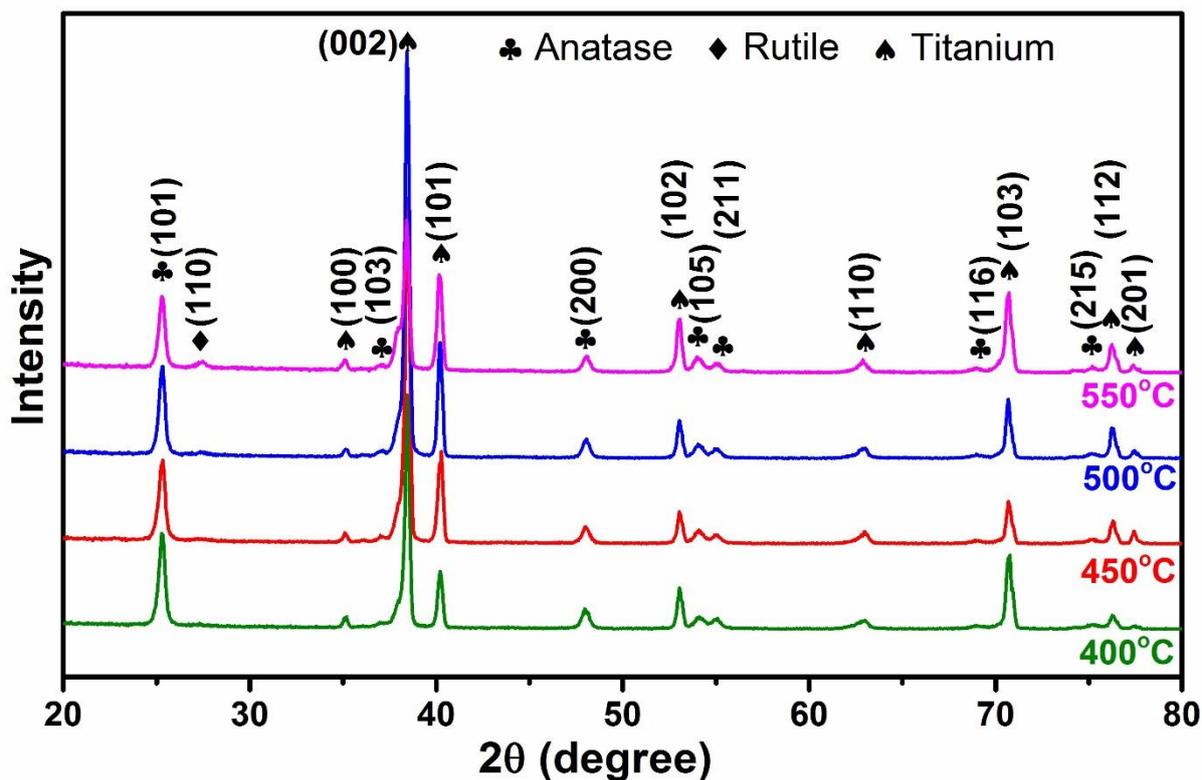

**Figure S10.** XRD patterns of TiO$_2$ nanotubes decorated with 1 nm-thick Au, 1 nm-thick Pt and 6 nm-thick Ag followed by a thermal dewetting process in Ar for 1 h at different temperatures: 400 °C, 450 °C, 500 °C and 550 °C. All samples were dealloyed in HNO$_3$ at 25 °C for 4 h.

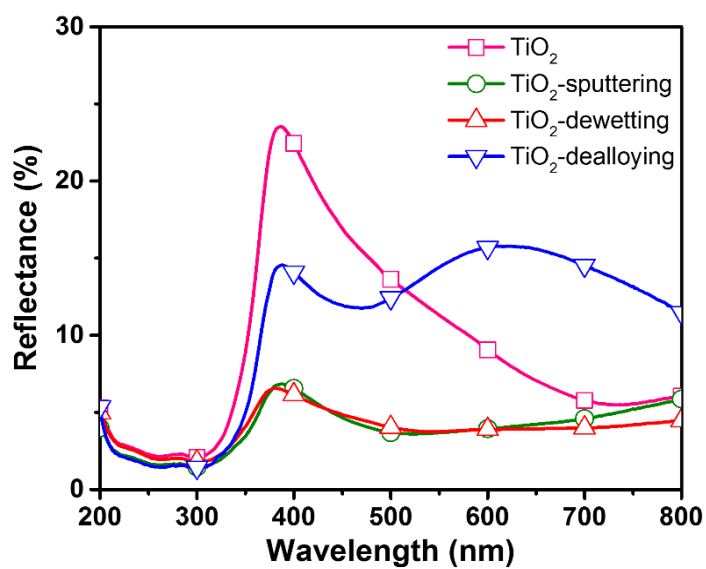

**Figure S11.** Reflectance measurements of TiO$_2$ NTs and Au$_1$Pt$_1$Ag$_6$-decorated TiO$_2$ NTs.

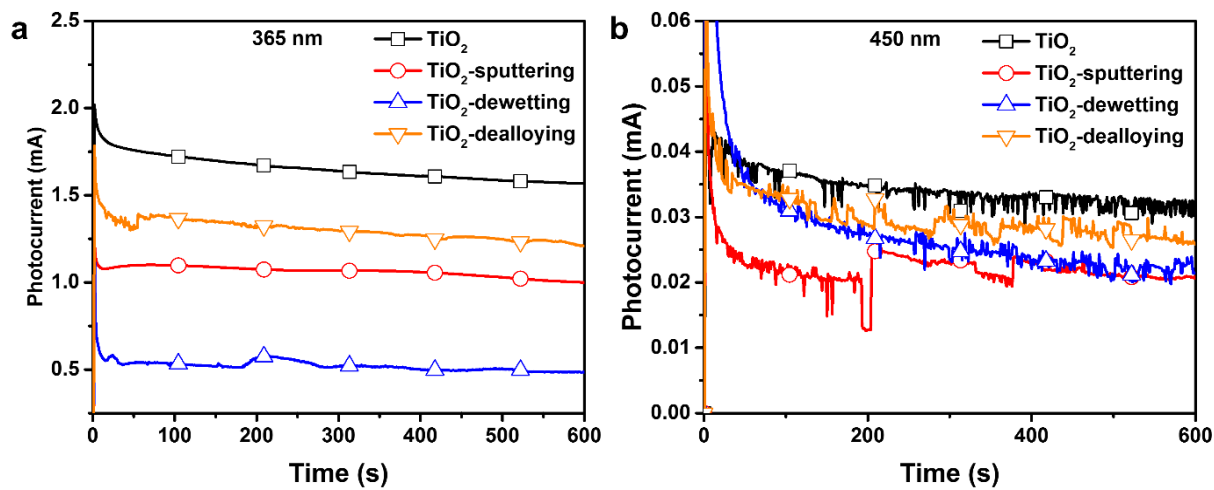

**Figure S12.** Photocurrent measurements of TiO$_2$ nanotubes and Au1Pt1Ag6-decorated TiO$_2$ NTs using a light source of: (a) 365 nm and (b) 450 nm.